\documentclass{ws-mpla}
\usepackage{amstext}
\usepackage{graphicx}
\usepackage{amssymb}

\usepackage{amsfonts}

\makeatother

\begin{document}

\title{On topological restrictions of the spacetime in cosmology}

\author{Torsten Asselmeyer-Maluga }

\address{German Aerospace Center, Berlin and Loyola University, New Orleans, LA, USA \\ torsten.asselmeyer-maluga@dlr.de}

\author{Jerzy Kr\'ol}

\address{University of Silesia, Institute of Physics, ul. Uniwesytecka 4,
40-007 Katowice, Poland \\ iriking@wp.pl}

\maketitle

\pub{Received (Day Month Year)}{Revised (Day Month Year)}

\begin{abstract}
In this paper we discuss the restrictions of the spacetime for the
standard model of cosmology by using results of the differential topology
of 3- and 4-manifolds. The smoothness of the cosmic evolution is the
strongest restriction. The Poincare model (dodecaeder model), the
Picard horn and the 3-torus are ruled out by the restrictions but
a sum of two Poincare spheres is allowed.
\end{abstract}

\ccode{PACS Nos.:04.60.Gw, 02.40.Ma, 04.60.Rt}



\section{Introduction}

In the 80's, there were a growing understanding of 3- and 4-manifolds.
Mike Freedman proved the (topological) Poincare conjecture in dimension
4 and classifies closed, compact, simple-connected, topological 4-manifolds
in 1982 \cite{Fre:82}. Bill Thurston \cite{Thu:97} presented its
geometrization conjecture at the same year (the geometrization conjecture
was proved by Perelman). Soon afterward, Simon Donaldson \cite{Don:83}
found a large class of non-smoothable closed, compact, simple-connected
4-manifolds leading to the first examples of exotic $\mathbb{R}^{4}$.
Beginning with this development, our understanding of 3- and 4-manifolds
as well its relation to each other is now in a better state. In physics,
4-manifolds are models for the spacetime and 3-manifolds are the spatial
part (like in global hyperbolic spacetimes $\Sigma\times\mathbb{R}$
with the 3-manifold $\Sigma$ as Cauchy surface). There are only few
papers \cite{Chernov2012} discussing the physical implications of
the new 3- and 4-dimensional results. But we do not know any paper
with a look for the cosmological implications.

\section{Preliminaries: 3- and 4-manifolds}

This section serves only as a short introduction into the theory of
3- and 4-manifolds. Further details can be found in the books \cite{Kir:89,FreQui:90,Thu:97,PrasSoss:97,GomSti:1999,Asselmeyer2007}.

\subsection{3-manifolds and geometric structures}

\begin{figure}
\includegraphics{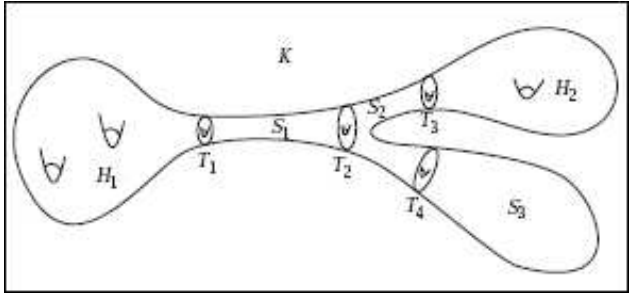}

\caption{Torus (JSJ-) decomposition, $H_{i}$ hyperbolic manifold, $S_{i}$
Graph-manifold, $T_{i}$ Tori \label{fig:Torus-decomposition}}

\end{figure}

A connected 3-manifold $N$ is prime if it cannot be obtained as a
connected sum of two manifolds $N_{1}\#N_{2}$ (see the appendix \ref{sec:Connected-and-boundary-connected}
for the definition) neither of which is the 3-sphere $S^{3}$ (or,
equivalently, neither of which is the homeomorphic to $N$). Examples
are the 3-torus $T^{3}$ and $S^{1}\times S^{2}$ but also the Poincare
sphere. According to \cite{Mil:62}, any compact, oriented 3-manifold
is the connected sum of an unique (up to homeomorphism) collection
of prime 3-manifolds (prime decomposition). A subset of prime manifolds
are the irreducible 3-manifolds. A connected 3-manifold is irreducible
if every differentiable submanifold $S$ homeomorphic to a sphere
$S^{2}$ bounds a subset $D$ (i.e. $\partial D=S$) which is homeomorphic
to the closed ball $D^{3}$. The only prime but reducible 3-manifold
is $S^{1}\times S^{2}$. For the geometric properties (to meet Thurstons
geometrization theorem) we need a finer decomposition induced by incompressible
tori. A properly embedded connected surface $S\subset N$ is called
2-sided%
\footnote{The 'sides' of $S$ then correspond
to the components of the complement of $S$ in a tubular neighborhood
$S\times[0,1]\subset N$.%
} if its normal bundle is trivial, and 1-sided if its normal bundle
is nontrivial. A 2-sided connected surface $S$ other than $S^{2}$
or $D^{2}$ is called incompressible if for each disk $D\subset N$
with $D\cap S=\partial D$ there is a disk $D'\subset S$ with $\partial D'=\partial D$.
The boundary of a 3-manifold is an incompressible surface. Most importantly,
the 3-sphere $S^{3}$, $S^{2}\times S^{1}$ and the 3-manifolds $S^{3}/\Gamma$
with $\Gamma\subset SO(4)$ a finite subgroup do not contain incompressible
surfaces. The class of 3-manifolds $S^{3}/\Gamma$ (the spherical
3-manifolds) include cases like the Poincare sphere ($\Gamma=I^{*}$
the binary icosaeder group) or lens spaces ($\Gamma=\mathbb{Z}_{p}$
the cyclic group). Let $K_{i}$ be irreducible 3-manifolds containing
incompressible surfaces then we can $N$ split into pieces (along
embedded $S^{2}$)\begin{equation}
N=K_{1}\#\cdots\#K_{n_{1}}\#_{n_{2}}S^{1}\times S^{2}\#_{n_{3}}S^{3}/\Gamma\,,\label{eq:prime-decomposition}\end{equation}
where $\#_{n}$ denotes the $n$-fold connected sum and $\Gamma\subset SO(4)$
is a finite subgroup. The decomposition of $N$ is unique up to the
order of the factors. The irreducible 3-manifolds $K_{1},\ldots,\, K_{n_{1}}$
are able to contain incompressible tori and one can split $K_{i}$
along the tori into simpler pieces $K=H\cup_{T^{2}}G$ \cite{JacSha:79}
(called the JSJ decomposition). The two classes $G$ and $H$ are
the graph manifold $G$ and hyperbolic 3-manifold $H$ (see Fig. \ref{fig:Torus-decomposition}).
The hyperbolic 3-manifold $H$ has a torus boundary $T^{2}=\partial H$,
i.e. $H$ admits a hyperbolic structure in the interior only. One
property of hyperbolic 3-manifolds is central: Mostow rigidity. As
shown by Mostow \cite{Mos:68}, every hyperbolic $n-$manifold $n>2$
has this property: \emph{Every diffeomorphism (especially every conformal
transformation) of a hyperbolic $n-$manifold is induced by an isometry.}
Therefore one cannot scale a hyperbolic 3-manifold and the volume
is a topological invariant. Together with the prime and JSJ decomposition\[
N=\left(H_{1}\cup_{T^{2}}G_{1}\right)\#\cdots\#\left(H_{n_{1}}\cup_{T^{2}}G_{n_{1}}\right)\#_{n_{2}}S^{1}\times S^{2}\#_{n_{3}}S^{3}/\Gamma\,,\]
we can discuss the geometric properties central to Thurstons geometrization
theorem: \emph{Every oriented closed prime 3-manifold can be cut along
tori (JSJ decomposition), so that the interior of each of the resulting
manifolds has a geometric structure with finite volume.} Now, we have
to clarify the term ''geometric structure''. A model geometry is a
simply connected smooth manifold $X$ together with a transitive action
of a Lie group $G$ on $X$ with compact stabilizers. A geometric
structure on a manifold $N$ is a diffeomorphism from $N$ to $X/\Gamma$
for some model geometry $X$, where $\Gamma$
is a discrete subgroup of $G$ acting freely on $X$. t is a surprising
fact that there are also a finite number of three-dimensional model
geometries, i.e. 8 geometries with the following models: spherical
$(S^{3},O_{4}(\mathbb{R}))$, Euclidean $(\mathbb{E}^{3},O_{3}(\mathbb{R})\ltimes\mathbb{R}^{3})$,
hyperbolic $(\mathbb{H}^{3},O_{1,3}(\mathbb{R})^{+})$, mixed spherical-Euclidean
$(S^{2}\times\mathbb{R},O_{3}(\mathbb{R})\times\mathbb{R}\times\mathbb{Z}_{2})$,
mixed hyperbolic-Euclidean $(\mathbb{H}^{2}\times\mathbb{R},O_{1,3}(\mathbb{R})^{+}\times\mathbb{R}\times\mathbb{Z}_{2})$
and 3 exceptional cases called $\tilde{SL}_{2}$ (twisted version
of $\mathbb{H}^{2}\times\mathbb{R}$), NIL (geometry of the Heisenberg
group as twisted version of $\mathbb{E}^{3}$), SOL (split extension
of $\mathbb{R}^{2}$ by $\mathbb{R}$, i.e. the Lie algebra of the
group of isometries of 2-dimensional Minkowski space). We refer to
\cite{Scott1983} for the details.

\subsection{4-manifolds and smoothness}

In this subsection we will shortly discuss the relation between 3-manifolds
and 4-manifolds. At first, every oriented, compact 3-manifold is the
boundary of a compact, simple-connected 4-manifold (Theorem 2 in chapter
VII of \cite{Kir:89}). Therefore we have to concentrate on simple-connected
4-manifolds, which are classified by Freedman \cite{Fre:82} topologically.
The topological classification based on the intersection form $\sigma(M)$
of a simple-connected, compact, closed 4-manifold. The intersection
form is a quadratic form over the second homology $H_{2}(M)$ (with
integer coefficients). An algebraic splitting of the form $\sigma(M)=\sigma_{1}\oplus\sigma_{2}$
is realized by a (topological) splitting of the 4-manifold (along
homology 3-spheres, see \cite{FreTay:77}). At this point, there is
a big difference between the smooth and the topological case: this
algebraic splitting of the form is not always realized by a smooth
splitting of the 4-manifold. A direct consequence is the fact that
all homology 3-spheres are bounding contractable, topological 4-manifolds
which are not always smoothable. An example is the Poincare sphere,
i.e. there is no contractable, smooth 4-manifold with boundary the
Poincare sphere. Therefore, the assumption of a smooth spacetime is
very restrictive as we will see below.

\section{The restrictions to smooth cosmological spacetimes}

According to the cosmological principle, our expanding universe, although
it is so complex, can be considered at very large scale homogeneous
and isotropic. The exact solution to Einstein's equation, describing
a homogeneous, isotropic universe, is in general called the Friedmann-Lemaitre-
Robertson-Walker (FLRW) metric. The FLRW model shows that the universe
should be, at a given moment of time, either in expansion, or in contraction.
From Hubble's observations, we know that the universe is currently
expanding. The FLRW-model shows that, long time ago, there was a very
high concentration of matter, which exploded in what we call the Big
Bang. The singularity theorems of Hawking and Penrose \cite{HawEll:94}
showed the necessity for the appearance of singularities including
the Big Bang under general circumstances. In this paper we will consider
the main hypothesis:\\
\textbf{Main Hypothesis}: \emph{Our universe is a compact 3-manifold
$\Sigma$ expanding smoothly so that the spacetime is a smooth 4-manifold
$M$. The topology of the 3-manifold is allowed to change. The spacetime
is a compact, smooth 4-manifold, i.e. we consider only the finite
time period from the Big Bang to the current universe so that $\partial M=\Sigma$.}\\
The compactness of the 3-manifold $\Sigma$ is a hypothesis motivated
by the WMAP data \cite{AurichLustigSteiner2007,WMAPcompactSpace2008,AurichLustigSteiner2008}.
One of the enigmas of the cosmic microwave background (CMB) is the
low power in the temperature correlations at large angles. Especially
the gaps in the spectrum of the background radiation have a simple
explanation by assuming a compact universe. Current further investigations
\cite{AurichLustig2011b} gave the conclusion that the low power at
large angles is real with high probability. One explanation of this
suppression of power could be that the universe possesses a non-trivial
topology (multi-connected spatial space i.e. compact 3-manifolds with
non-trivial fundamental group). Currently three models are discussed,
the Poincare (or dodecaeder) model \cite{dodecaeder:03} (positive
curvature), the Picard horn \cite{AurichLustigSteinerThen2004} (negative
curvature) and the flat universe with the 3-torus (homogeneous model
\cite{AurichLustigSteiner2007,AurichLustigSteiner2008}) or the half-turn
space (inhomogeneous model \cite{AurichLustig2011}). But the current
data are unable to decide between these cases \cite{AurichLustig2011b}.

Now we study the implications of this hypothesis by using some results
of the differential topology of 3- and 4-manifolds.

\subsection{Classical case}

In the classical case we assume a singularity, i.e. the existence
of a point in the past attracting all geodesics (pointed backward).
A simple example is given by a cone over the circle. All normals to
the circle (the geodesics of the cone) converge to the apex. At the
same time one can interpret the cone as a process to contract the
circle to a point. Then one uses these geodesics to construct this
(continuous) contraction map. To generalize this case, we remark that
the cone is homeomorphic (actually also diffeomorphic after smoothing
the apex) to the disk $D^{2}$. Therefore we take the 4-disk $D^{4}$
homeomorphic to the cone $Cone(S^{3})\simeq D^{4}$ over $S^{3}$
($\simeq$ denotes ''homeomorphic''). This 4-manifold (compact with
boundary) serves as a model for the ''classical'' Big Bang, starting
with a singularity (the apex) and evolving to the 3-sphere after a
finite time. Topologically, the model can be characterized by the
property: the cone $Cone(S^{3})$ is contractable to a point, i.e.
there is a continuous (actually also a smooth) homotopy $Cone(S^{3})\to\left\{ \star\right\} $.
But we may ask: Is this model the most general case? We will partly
answer this question in the following.

To fix the problem, we have to consider the class of smooth, compact,
contractable 4-manifolds. Contractability is needed (but not necessary)
to obtain a Lorentz metric outside of the Big Bang singularity%
\footnote{Remember, at the singularity all geodesics (with backward time orientability)
converges. To see it, consider a non-vanishing vector field which
is the obstruction to introduce a Lorentz metric \cite{Ste:99}. For
compact manifolds, one needs a vanishing Euler characteristics (Poincare-Hopf
theorem). A contractable manifold $A$ has Euler characteristics $\chi(A)=1$
but the excision of one point gives the desired result $\chi(A\setminus\{\star\})=0$.%
}. Furthermore contractability guarantees that there is no topology
change after the Big Bang. We will later discuss the case of an explicit
topology change. We remark that in topology there exists also a general
procedure to construct the cone $cone(\Sigma$) over a topological
space $\Sigma$ by\[
cone(\Sigma)=\frac{\Sigma\times[0,1]}{\Sigma\times\{1\}\sim\{\star\}}\]
where $\Sigma\times\left\{ 1\right\} $ is contracted to a point $\left\{ \star\right\} $
(the apex). The cone $cone(\Sigma$) is contractable but it is in
most cases not a manifold but an orbifold (it fails to be locally
modeled on open subsets of $\mathbb{R}^{4}$ but instead on quotients
of open subsets of $\mathbb{R}^{4}$ by finite group actions). Therefore
a whole neighborhood of the singularity fails to be a manifold. 
To illustrate this singularity we consider the cone over the 3-sphere
again. The homeomorphism between the cone $Cone(S^{3})$ and the 4-disk
is known as Alexanders trick, i.e. a homeomorphism of the two boundaries
$\partial Cone(S^{3})=S^{3}$ and $\partial D^{4}=S^{3}$ can be extended
to a homeomorphism $Cone(S^{3})\simeq D^{4}$. This fails for the
3-torus. The cone $Cone(T^{3})$ contains a 3-torus for every value
$t\in[0,1)$ but not at $t=1$ (where it is a point). Therefore the
neighborhood of this point do not look like $T^{3}\times[a,b]$ (with
the finite interval $[a,b]$), $t=1$ is a so-called non-flat point
in topology. Especially the neighborhood is not an open subset of
$\mathbb{R}^{4}$ (but rather $\mathbb{R}^{4}/\mathbb{Z}_{3}$). All
tangent vectors at $t=1$ vanishes, i.e. the point is non-smooth and
cannot be smoothly approximated (like in the case $Cone(S^{3})$).
The whole discussion remains true for every 4-space $Cone(\Sigma)$
with $\Sigma$ a 3-manifold having a different homology then the 3-sphere
(see below).
Our current understanding of the singularity (see \cite{Stoica2011})
gives a manifold structure in the neighborhood of the singularity,
i.e. Einsteins equation gives a smooth solution except for the singular
point. Therefore the model of the 3-torus, the Picard horn or half-turn
space is singled out (or one has to assume a non-smooth transition).
But then we are left with a contractable, smooth 4-manifold as a model
of the spacetime (for finite times). 

\begin{figure}
\includegraphics[scale=0.5]{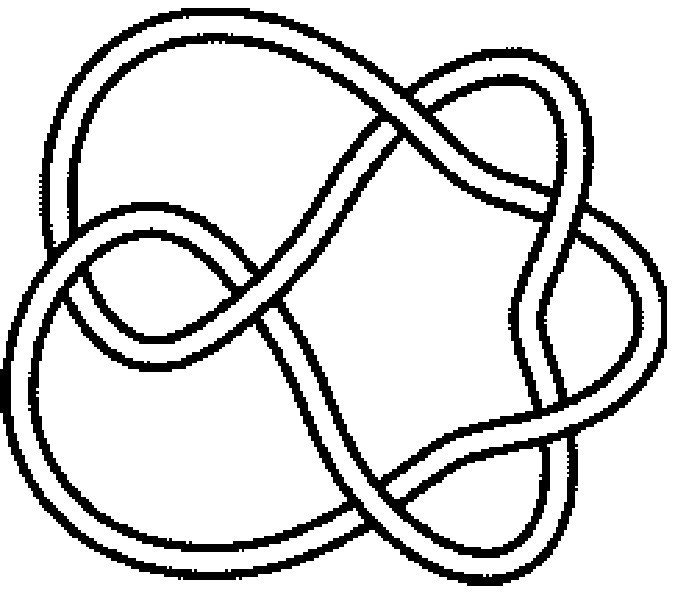} \includegraphics[scale=0.5]{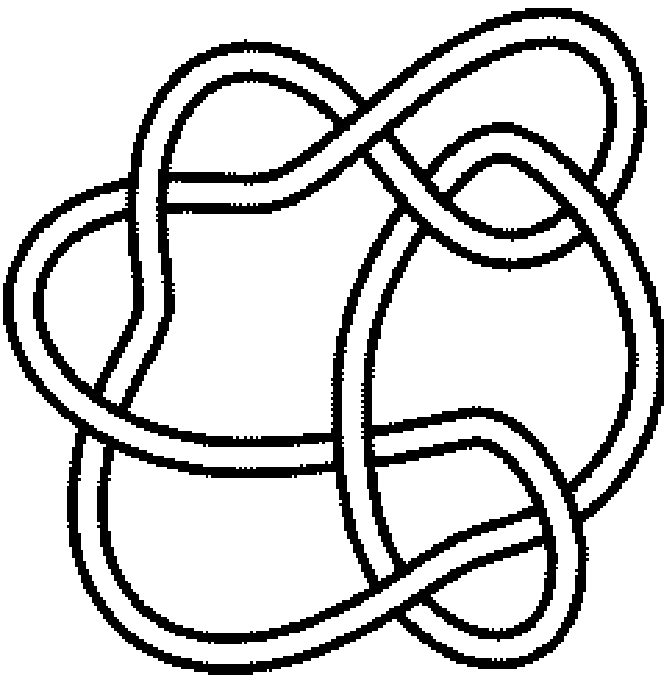}
\includegraphics[scale=0.5]{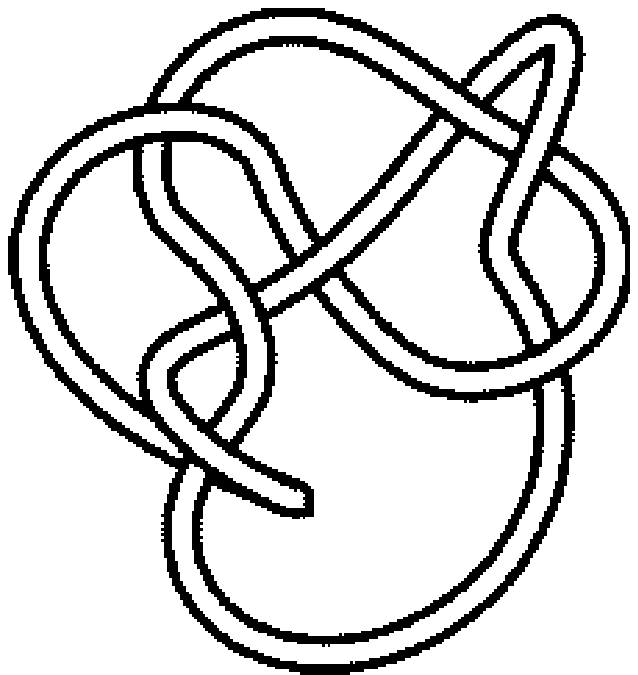}

\caption{$\pm1$ Dehn surgery along these knots ($6_{1},8_{8},8_{9}$) generates
3-manifolds bounding smoothly contractable 4-manifolds\label{fig:-Dehn-surgery-contractable-4MF}}

\end{figure}

\begin{figure}
\includegraphics[scale=0.5]{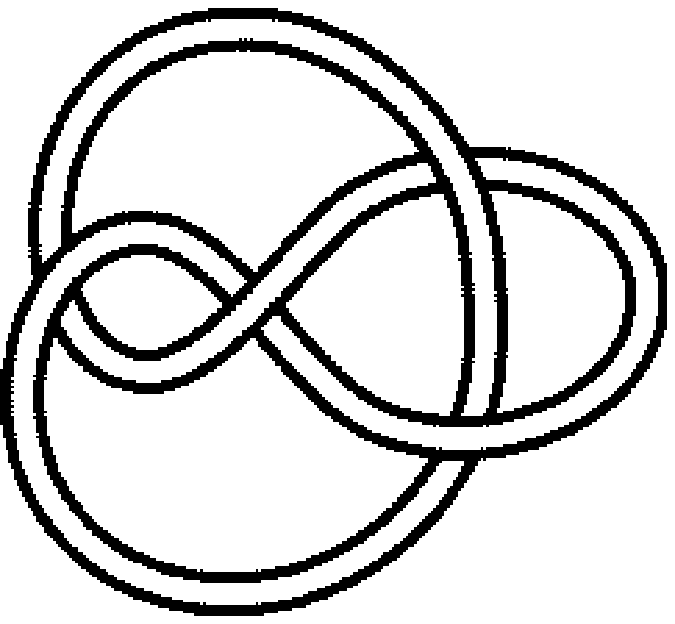} \includegraphics[scale=0.5]{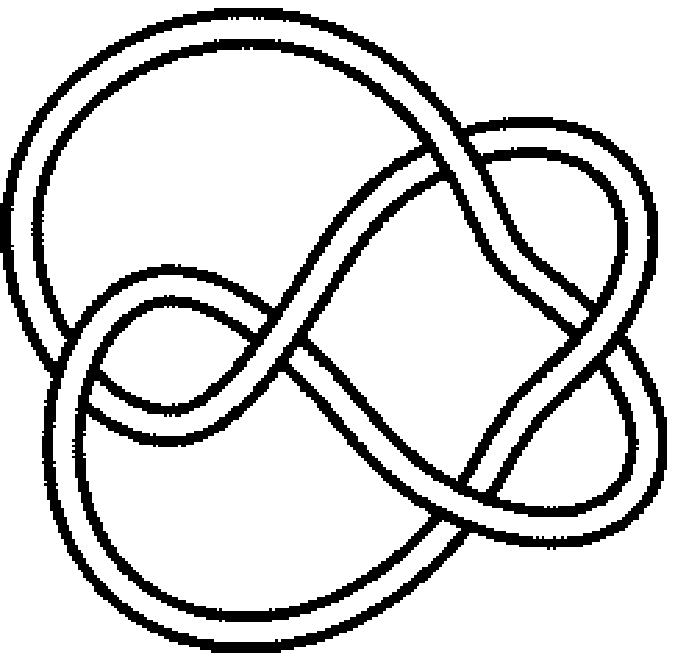}
\includegraphics[scale=0.5]{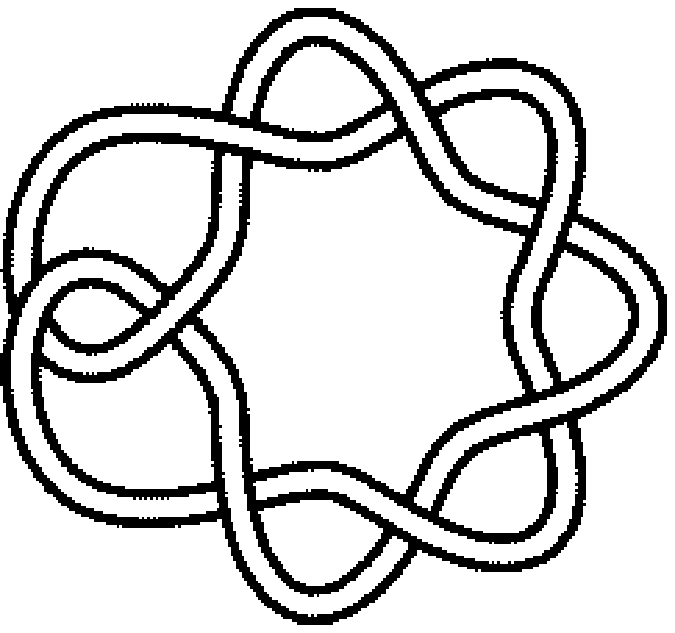}

\caption{$\pm1$ Dehn surgery along these knots generates $(4_{1},5_{2},8_{1})$
3-manifolds bounding only non-smoothly contractable 4-manifolds\label{fig:-Dehn-surgery-non-contractable-4mf}}

\end{figure}

Freedman \cite{Fre:79,Fre:82} solves the problem to find all compact,
contractable, topological 4-manifolds. He proved that any homology
3-sphere (i.e. a compact 3-manifold with the same homology groups
as the 3-sphere) bounds a compact, contractable, topological 4-manifold.
But as a corollary to Donaldson's theory \cite{Don:83,Don:86,Don:87},
not all homology 3-spheres are the boundary of a \textbf{smooth} compact
contractable 4-manifold (see also \cite{FinSte:85}). One example
is the Poincare homology 3-sphere bounding only a topological (but
non-smooth) compact contractable 4-manifold. A further division for
all possible cases is given by the inclusion of the geometric structure,
i.e the existence of homogeneous metrics on $\Sigma$. Especially
we concentrated our discussion to the case of prime 3-manifolds. The
following cases for prime homology 3-spheres are possible:
\begin{enumerate}
\item Spherical geometry (positive scalar curvature): There are only two
homology 3-spheres with this geometry: the 3-sphere $S^{3}$ and the
Poincare sphere (or sums of these cases). The Poincare sphere can
be excluded by Donaldson theory (see above). The geometry $S^{2}\times\mathbb{R}$
can be also excluded, there is no homology 3-sphere with this geometry.
\item Flat geometry (Euclidean geometry or NIL geometry): There are no homology
3-sphere carrying a homogeneous, flat metric. 
\item Negative curvatures (hyperbolic and $\tilde{SL}_{2}$ geometry): The
class of homogeneous metrics with a negative curvature (at least along
one direction) forms the largest class of possible geometric structures
on homology 3-spheres. All members of the special class of Seifert
fibred homology 3-spheres have the $\tilde{SL}_{2}$ geometry. An
example is the Brieskorn sphere $\Sigma(2,5,7)$ defined by the set
\[
\Sigma(p,q,r)=\left\{ (u,v,w)\in\mathbb{C}^{3}\,|\: u^{p}+v^{q}+r^{r}=0\right\} \cup S^{5}\]
for $p=2,q=5,r=7$ (which is the boundary of the famous Mazur manifold
\cite{AkbKir:79}, a contractable smooth 4-manifold). Examples of
hyperbolic homology 3-spheres can be constructed by using $\pm1$
Dehn surgery \cite{Rol:76} along hyperbolic knots%
\footnote{A knot $K$ as smooth embedding $S^{1}\to S^{3}$ is hyperbolic if
its complement $S^{3}\setminus\left(K\times D^{2}\right)$ is a hyperbolic
3-manifold (with boundary a torus), i.e. a 3-manifold with a homogeneous
metric in the interior. %
}. Examples of hyperbolic 3-manifolds bounding contractable smooth
4-manifolds are generated by the knots%
\footnote{These knots are all members of the class of slice knots, i.e. knots
bounding a disk.%
} $6_{1}$ or $8_{8}$ (in Rolfsen notation, see Fig. \ref{fig:-Dehn-surgery-contractable-4MF})
whereas hyperbolic knots like the figure 8-knot $4_{1}$ or the knot
$5_{2}$ are excluded (see Fig. \ref{fig:-Dehn-surgery-non-contractable-4mf}).
\end{enumerate}
Now we may expect that the understanding of the simplest pieces is
enough to get an overview for the general case. But as usual, 4-dimensional
topology is an exception. The connected sum of prime homology 3-spheres
can bound a contractable, compact, smooth 4-manifold. As we saw above
the Poincare sphere $P$ cannot be a boundary of this kind but the
following sum\[
P\#P\#K_{1}\#K_{2}\#K_{3}\]
where $K_{i}$ are prime hyperbolic homology 3-spheres%
\footnote{There is only one condition: $K_{i}$ has to smoothly embed into $S^{2}\times S^{2}$.%
} (for instance generated by the knots $6_{1}$, $8_{8}$ or $8_{9}$
via $\pm1$ Dehn surgery, see Fig. \ref{fig:-Dehn-surgery-contractable-4MF}).
Therefore, if the Poincare model \cite{dodecaeder:03} of the cosmos
is correct then one has to use the sum of two Poincare spheres. Otherwise
one has to assume a non-smooth topology change in the primordial phase
of the cosmos.

\subsection{Big bounce case}

Our discussion has excluded (like the 3-torus) or restricted (like
the Poincare sphere) many interesting models of the cosmos. Here we
want to consider the possible restrictions for a model without singularity
inspired by Loop quantum gravity. In \cite{AshPawSin:06}, Ashtekar
et.al. described a cosmological model showing the effect of a Big
bounce (see also \cite{AshtekarLQC2011}). Then this model does not
show a singularity, i.e. there is no Big Crunch and the contraction
is followed by an expansion again. The crucial point is the determination
of the initial state. The following assumptions are physically plausible:
\begin{itemize}
\item The (spatial) cosmos at the ''Big Bounce point'' is closed (no boundary).
\item It consists of a minimal number of elementary cells.
\end{itemize}
The first assumption favors a compact, closed 3-manifold. Using homology
theory, we have to use at least two cells (glued together along the
common boundary) to fulfill the second assumption. It singles out
the 3-sphere (uniquely using the solution to the Poincare conjecture).
Therefore the Big Bounce model starts with a (Planck-sized) 3-sphere
$S^{3}$ and evolved smoothly to a closed, compact 3-manifold $\Sigma$(see
the Main hypothesis above). Although this starting point is different
from the singularity in the previous subsection, the results are quite
similar.

At first we remark that the evolution from the 3-sphere $S^{3}$ to
$\Sigma$ is a smooth 4-manifold, a cobordism $W(S^{3},\Sigma)$ between
$S^{3}$ and $\Sigma$. We can close partly this cobordism along the
3-sphere by using a 4-disk, $W(S^{3},\Sigma)\cup_{S^{3}}D^{4}$, to
get a contractable 4-manifold with boundary $\Sigma$. Therefore we
obtain the same restrictions as above: $\Sigma$ must be a homology
3-sphere which bounds a smooth, contractable, compact 4-manifold.
Again the case of a single Poincare sphere is excluded (by Donaldson
theory). The list of prime homology 3-spheres agrees with the list
above. 

A difference to the results above is the flat case (i.e. with Euclidean
or NIL geometry). The simplest example is the 3-torus $T^{3}$ \cite{AurichLustigSteiner2008}.
Now we have to consider a (smooth) cobordism $W(S^{3},T^{3})$ between
the 3-sphere and the 3-torus. But then we obtain a topology change
from the simple-connected 3-sphere to the multiple-connected 3-torus.
Especially the cobordism $W(S^{3},T^{3})$ itself is not a simple-connected,
compact 4-manifold (in contrast to a cobordism $W(S^{3},\Sigma)$
between $S^{3}$ and the homology 3-sphere). As discussed in \cite{Dowker1997},
the cobordism $W(S^{3},T^{3})$ (a Morse cobordism in the notation
of \cite{Dowker1997}) gives rise to causal discontinuities (in the
sense of the Borde-Sorkin conjecture \cite{Dowker1997,BordeDowkerSorkin1999})
and allows the singular propagation of a quantum field (like in the
trousers spacetime in $1+1$ dimensions \cite{deWitteAnderson1986}).
But there is also a second argument against the appearance of the
cobordism $W(S^{3},T^{3})$: Clearly $W(S^{3},T^{3})$ is a Lorentz
cobordism but the mod-2 Kervaire semi-characteristics $\mathcal{U}(\partial W(S^{3},T^{3}))$
gives\begin{eqnarray*}
\mathcal{U}(\partial W(S^{3},T^{3})) & = & \left(\dim H^{0}(\partial W(S^{3},T^{3}),\mathbb{Z}_{2})+\dim H^{1}(\partial W(S^{3},T^{3}),\mathbb{Z}_{2})\right)\bmod2\\
 & = & 5\bmod2=1\end{eqnarray*}
i.e. the cobordism $W(S^{3},T^{3})$ is not a Spin-Lorentz cobordism
(see \cite{GibbonsHawking1992,Giulini1992}) or it do not admit a
$SL(2,\mathbb{C})-$spin structure (defining an unique parallel transport
of a spinor). The case of a NIL geometry is similar. The corresponding
3-manifold $\tilde{T}^{3}$ is a twisted 3-torus%
\footnote{This twisted 3-torus is the mapping torus $M_{f}(T^{2})$: Consider
$T^{2}\times[0,1]$ and identify $T^{2}\times\left\{ 0\right\} $
and $T^{2}\times\left\{ 1\right\} $ by a diffeomorphism $f:T^{2}\to T^{2}$.
If $f$ is of finite order then $M_{f}(T^{2})$ has Euclidean geometry
and if $f$ is a Dehn twist then $M_{f}(T^{2})$ has NIL geometry.
The remaining case (Anosov map) leads to the SOL geometry (as twisted
$\mathbb{H}^{2}\times\mathbb{R}$ geometry with negative curvature).%
} but leading to the same mod-2 Kervaire semi-characteristics $\mathcal{U}(\partial W(S^{3},\tilde{T}^{3}))=1$
with the same result: it is not a Spin-Lorentz cobordism.

\section{Conclusion}

In this paper we discussed the differential-topological restrictions
on the spacetime for the evolution of our universe with an explicit
Big Bang singularity and for models with the Big Bounce effect. Surprisingly,
the results are the same for both cases. Furthermore, we showed that
the main restriction comes from the assumption of a smooth spacetime.
The relaxation of this smoothness assumption causes in bad singularities
like non-manifold points or complicated topology changes with causal
discontinuities. Especially the Poincare sphere, the Picard horn or
the 3-torus are not favored.

\section*{Acknowledgments}
This work was partly supported (T.A.) by the LASPACE grant. Many
thanks to Carl H. Brans for all the discussions about the physics
of exotic 4-manifolds. The authors acknowledged for all mathematical
discussions with Duane Randall, Bob Gompf and Terry Lawson. 

\appendix

\section{Connected and boundary-connected sum of manifolds\label{sec:Connected-and-boundary-connected}}

Now we will define the connected sum $\#$ and the boundary connected
sum $\natural$ of manifolds. Let $M,N$ be two $n$-manifolds with
boundaries $\partial M,\partial N$. The \emph{connected sum} $M\#N$
is the procedure of cutting out a disk $D^{n}$ from the interior
$int(M)\setminus D^{n}$ and $int(N)\setminus D^{n}$ with the boundaries
$S^{n-1}\sqcup\partial M$ and $S^{n-1}\sqcup\partial N$, respectively,
and gluing them together along the common boundary component $S^{n-1}$.
The boundary $\partial(M\#N)=\partial M\sqcup\partial N$ is the disjoint
sum of the boundaries $\partial M,\partial N$. The \emph{boundary
connected sum} $M\natural N$ is the procedure of cutting out a disk
$D^{n-1}$ from the boundary $\partial M\setminus D^{n-1}$ and $\partial N\setminus D^{n-1}$
and gluing them together along $S^{n-2}$ of the boundary. Then the
boundary of this sum $M\natural N$ is the connected sum $\partial(M\natural N)=\partial M\#\partial N$
of the boundaries $\partial M,\partial N$.

\section*{References}


\end{document}